\documentclass{aastex631}

\shorttitle{Geart storms estimated by Burton, O'Brien and McPherron equations}
\shortauthors{Zhao et al.}
\graphicspath{{./}{figures/}}

\begin{document}

\title{Can we estimate the intensities of great geomagnetic storms($\Delta$SYM-H$\le -$200 nT) by Burton equation or by O'Brien and McPherron equation?}

\correspondingauthor{Gui-Ming Le}
\email{legm@cma.gov.cn}

\author[0000-0002-1031-018X]{Ming-Xian Zhao}
\affiliation{Key Laboratory of Space Weather, National Center for Space Weather \\
China Meteorological Administration \\
Beijing, 100081, China}
\affiliation{Key Laboratory of Solar Activity, National Astronomical Observatories \\
Chinese Academy of Sciences \\
Beijing, 100012, China}

\author[0000-0002-9906-5132]{Gui-Ming Le}
\affiliation{Key Laboratory of Space Weather, National Center for Space Weather \\
China Meteorological Administration \\
Beijing, 100081, China}
\affiliation{Key Laboratory of Solar Activity, National Astronomical Observatories \\
Chinese Academy of Sciences \\
Beijing, 100012, China}

\author[0000-0001-7042-5395]{Jianyong Lu}
\affiliation{Institute of Space Weather, School of Math \& Statistics \\
 Nanjing University of Information Science \& Technology \\
 Nanjing, China}

\begin{abstract}

We input solar wind parameters responsible for the main phases of 15 great geomagnetic storms (GGSs: $\Delta$SYM-H$\le-$200 nT) into the empirical formulae created by \cite{Burton1975}(hereafter Burton equation), and by \cite{OBrien2000}(hereafter OM equation) to evaluate whether \textbf{two equations} can correctly estimate the intensities of GGSs. The results show that the intensities of most GGSs estimated by OM equation are much smaller than the observed intensities. The RMS error between the intensities estimated by OM equation and the observed intensities is \textbf{203} nT, implying that the estimated storm intensity deviates significantly from the observed one. The RMS error between the intensities estimated by Burton equation and the observed intensities is 130.8 nT. The relative error caused by Burton equation for the storms with intensities $\Delta$SYM-H$<$-400 nT is larger than 27\%, implying that the absolute error will be large for the storms with $\Delta$SYM-H$<$-400 nT. \textbf{The results indicate that the two equations cannot work effectively in the estimation of GGSs. On the contrary, the intensity of a GGS estimated by the empirical formula created by \cite{WangCB2003} can always be very close to the observed one if we select the right weight for solar wind dynamic pressure, proving that solar wind dynamic pressure is an important factor for  GGS intensity, but it is overlooked in the ring current injection terms of Burton equation or OM equation. This is the reason why the two equations cannot work effectively in the estimation of GGSs.}

\end{abstract}

\keywords{Solar wind(1534) --- Solar-terrestrial interactions(1473) --- Space weather(2037)}


\section{Introduction} \label{sec-intro}

A geomagnetic storm is a worldwide intense disturbance of the geomagnetic field. Various solar wind structures such as sheath between an interplanetary shock and the corresponding interplanetary coronal mass ejection (ICME), ICME, corotating interaction region (CIR) \citep{Dal2006}, which is caused by the interaction between high speed solar wind from a coronal hole and low speed solar wind, or the combination of these structures, can trigger a geomagnetic storm. A geomagnetic storm only caused by a sheath, an ICME or the combination of a sheath and an ICME is named as a CME-driven storm. The ICME within which the magnetic field changes gradually and the proton $\beta$ is low \citep{Zurbuchen2006} is called magnetic cloud (MC). Occasionally, a storm may be caused by both an ICME and a CIR \citep{Dal2006}. However, in the vast majority of cases, a geomagnetic storm is usually either a CME-driven storm or CIR-driven storm. \cite{Yermolaev2012} studied the geoeffectiveness and the efficiency of various solar wind structures and found that \textbf{CIR and sheath structures have higher occurrence and higher efficiency to produce geomagnetic storms than MCs. However, MCs usually have larger and longer duration of southward interplanetary magnetic field than CIR and sheath, they play an essential role in the generation of geomagnetic storms}. If we use incorrect criteria to identify the interplanetary drivers for geomagnetic storms, we will get incorrect results, where the contributions to geomagnetic storms made by sheath compression regions are not taken into account and the sheath role is often underestimated \citep{Yermolaev2021}.

It has found that most of major geomagnetic storms (Dst $\le -$100 nT) were CME-driven storms, while storms with Dst $> -$100 nT can be both CME-driven storms and CIR-driven storms \citep{Gosling1991, Richardson2002}. Differences between CME-driven storms and CIR-driven storms have been studied by \cite{Borovsky2006}. Some CME-driven storms were accompanied with solar proton events (SPEs). For example, about 80\% of the great geomagnetic storms (Dst $\le -$200 nT) during solar cycles 23 were accompanied with SPEs \citep{Le2016}. In this context, CME-driven storms pose more threaten to the instruments on spaceships, astronauts in space stations, and Earth-based power grid systems. \textbf{Many articles were devoted to the study of the relationship between CIR-driven storms and the environment of electrons of radiation belt \citep[e.g.,][and reference therein]{Borovsky2006, Shen2017, Pandya2019}. CIR-driven storms are more  effect for the accelation of electrons in the region with L $>$ 3.5 than CME-driven storms \citep{Miyoshi2005}. The longer duration of CIR-driven storms, would accelerate electrons to higher energies, cause stronger spacecraft charging, and produce higher fluxes of relativistic electrons in the outer radiation belt \citep{Borovsky2006}.} The events of high fluxes of relativistic electrons in geostationary orbit did not appear around solar maximum \citep{Le2021Sol2961}. However, most of GLEs, strong SPEs (peak flux $\ge$1000 pfu), and major geomagnetic storms appeared around solar maximum \citep{Le2020Sol2952, Le2021MNRAS, Le2021Solinpress}, indicating that solar sources for high fluxes of relativistic electrons in geostationary orbit are different from those for GLEs, strong SPEs and major geomagnetic storms.

The SYM-H index has a high time resolution (1 min) \citep{Iyemori1990, Iyemori2010} and can be treated as high time resolution of Dst index \citep{Wanliss2006}. The SYM-H index has been widely used in the study of the relationship between the solar wind and magnetosphere \citep[e.g.,][]{Bhaskar2019, Grandin2019, Hajra2020, Arowolo2021}. The response processes of the magnetosphere to different interplanetary disturbances are very complicated \citep[e.g.,][]{Kozyra2003, Du2008}. Various couple functions between the solar wind and the magnetosphere were proposed \citep[e.g.,][]{Kan1979, Akasofu1981, Wygant1983, Newell2007, Tenfjord2013} and \cite{Borovsky2014} made the canonical correlation analysis of the combined solar-wind and geomagnetic-index data. However, the empirical formula relating the geomagnetic storm intensity to the solar wind parameters based on these coupling functions has not been built. So far, some empirical formulae relating the geomagnetic storm intensity to the solar wind parameters were established \citep [e.g.,][]{Burton1975, Fenrich1998, OBrien2000, Temerin2002, Ballatore2003, WangCB2003, WangYM2003, Temerin2006, Boynton2011}. Of these empirical formulae, the empirical formula created by \cite{Burton1975}(hereafter Burton equation) and the empirical formula created by \cite{OBrien2000}(hereafter OM equation) were used most frequently compared with other formulae. The injection term of the ring current of a geomagnetic storm in Burton equation is a linear function of the dawn-to-dusk electric field of the solar wind (hereafter solar wind electric field). \cite{OBrien2000} modified the Burton equation and built a revised empirical formula. However, the injection term of the ring current in OM equation is still a linear function of the solar wind electric field. \cite{WangCB2003} built a empirical formula to describe the relationship between the geomagnetic storm intensity and the corresponding solar wind parameters (Hereafter WCL equation). In WCL equation, the injection term of the ring current is not only a function of the solar wind electric field, but also the function of the solar wind dynamic pressure.

The intensity of a geomagnetic storm is the result of the sustained interaction between solar wind and the magnetosphere during the main phase of a storm. The most outstanding advantage of empirical formulae is its simplicity in space weather forecast: by inputting the solar wind parameters responsible for the main phase of a geomagnetic storm into an empirical formula, one can quickly get the estimated intensity of this geomagnetic storm. Of the empirical formulae, Burton equation and OM equation were used mostly to predict the extreme geomagnetic storms. For example, Burton equation was used to estimate the intensity of the storm that occurred on September 1-2, 1859 by \cite{Tsurutani2003}, and \cite{Liu2014} used the OM equation to estimate the intensity of the storm on 24 July 2014. Extreme geomagnetic storms can cause widespread interference and damage to technological systems \citep[][and reference therein]{Love2021} and then lead to significant economic loss \citep[e.g.,][]{Baker2008, Schulte2014, Eastwood2017, Riley2017}. Hence, it is very important to ensure the forecast accuracy of extreme geomagnetic storms. Now the question is whether Burton equation or OM equation can estimate the intensities of very large geomagnetic storms correctly?  To answer this question, we compare the performance of three models mentioned above using 15 great geomagnetic storms ($\Delta$SYM-H$\le -$200 nT) that occurred during solar cycle 23. The solar wind parameters responsible for the main phases of 15 storms are inputted into three models to get the estimated intensities of the great geomagnetic storms (GGSs), which are compared with the observed intensities. To avoid any possible confusions, we use $\Delta$SYM-H$_b$, $\Delta$SYM-H$_{om}$, and $\Delta$SYM-H$_w$ to indicate the intensities of a geomagnetic storm estimated by Burton equation, OM equation, and WCL equation, respectively, while we use $\Delta$SYM-H$_{ob}$ to represent the observed intensity, namely the real variation of SYM-H index during the main phase of a geomagnetic storm. The rest part of the article is organized as follows. Data source and method are presented in Section \ref{sec-data}, results are presented in Section \ref{sec-results}, discussion and summary are presented in Section \ref{sec-discussion}.

\section{Data and Method} \label{sec-data}

\subsection{Data} \label{subsec-data}

The solar wind data with time resolution one minute are used in this study and are available from the website at \url{https://omniweb.gsfc.nasa.gov/form/omni_min.html}. The SYM-H index, which has the time resolution 1 min, is used to describe the intensity of a geomagnetic storm, can be obtained from the website  \url{http://wdc.kugi.kyoto-u.ac.jp/aeasy/index.html}.

\subsection{Method} \label{subsec-method}

The three empirical formulae can be written as below
\begin{equation}\label{eq-01}
  d \textrm{SYM-H}^*_{estimated}/d t = Q - D
\end{equation}
where $Q$ and $D$ are the injection term and decay term of the ring current of the associated geomagnetic storm, respectively. SYM-H$_{estimated}^*$ is the pressure corrected storm intensity estimated by the corresponding empirical formula. If we use Burton equation to estimate the intensity of a geomagnetic storm, SYM-H$_{estimated}^*$ will be defined as SYM-H$_{b}^*$, and $Q$ and $D$ will be $Q_b$ and $D_b$, respectively. Similarly, for OM method, SYM-H$_{estimated}^*$ is SYM-H$_{om}^*$, and $Q$ and $D$ are $Q_{om}$ and $D_{om}$, respectively, and for WCL approach, SYM-H$_{estimated}^*$ is SYM-H$_{w}^*$, and $Q$ and $D$ are $Q_{w}$ and $D_{w}$, respectively.

According to equation \ref{eq-01}, we obtain the following expression
\begin{equation} \label{eq-02}
\Delta \textrm{SYM-H}_{estimated}^* = \int_{t_s}^{t_e}(Q-D)d t = \int_{t_s}^{t_e}Q d t - \int_{t_s}^{t_e}D d t
\end{equation}
where $t_s$ and $t_e$ are the start and end times of a storm main phase, respectively. $\Delta$SYM-H$_{estimated}^*$is the variation of SYM-H$_{estimated}^*$ during the main phase of the storm.

We set $I(Q)$ and $I(D)$ as the time integral of $Q$ and $D$ during the main phase of a geomagnetic storm, respectively, then we have
\begin{equation}\label{eq-03}
\Delta \textrm{SYM-H}_{estimated}^* = I(Q) - I(D)
\end{equation}
SYM-H$_{ob}^*$, the pressure corrected SYM-H$_{ob}$, is calculated as follows according to OM equation,
\begin{equation}\label{eq-04}
\textrm{SYM-H}_{ob}^* = \textrm{SYM-H}_{ob} - b \sqrt{P_d} + c
\end{equation}
\begin{equation}\label{eq-05}
\Delta\textrm{SYM-H}_{ob}^* = \textrm{SYM-H}_{ob}^*(t_e) - \textrm{SYM-H}_{ob}^*(t_s)
\end{equation}
where SYM-H$_{ob}^*(t_e)$ and SYM-H$_{ob}^*(t_s)$ are the values of SYM-H$_{ob}^*$ at the end and start times of the main phase of a geomagnetic storm, respectively. $P_d$ is the solar wind dynamic pressure. The b and c were 7.26 and 11, respectively, in the article by \cite{Sandhu2021}, which are also used in the present study.

Since
\begin{equation}\label{eq-06}
  \left\{ \begin{array}{ll}
                \textrm{SYM-H}_{ob}^*(t_s) = \textrm{SYM-H}_{ob}(t_s) - 7.26 \sqrt{P_d}|_{t_s} + 11, \\
				\textrm{SYM-H}_{ob}^*(t_e) = \textrm{SYM-H}_{ob}(t_e) - 7.26 \sqrt{P_d}|_{t_e} + 11,
          \end{array} \right.
\end{equation}
We have
\begin{equation}\label{eq-07}
\Delta \textrm{SYM-H}_{ob}^* = \textrm{SYM-H}_{ob}(t_e)-\textrm{SYM-H}_{ob}(t_s)+7.26 \left( \sqrt{P_d}|_{t_s} - \sqrt{P_d}|_{t_e} \right)
\end{equation}
Where $\sqrt{P_d}|_{t_e}$ and $\sqrt{P_d}|_{t_s}$ are the values of $\sqrt{P_d}$ at the end and start times of the main phase of a geomagnetic storm, respectively.

Note that when we use WCL equation to estimate the intensity of a geomagnetic storm, we use the following expression as the injection term,
\begin{equation}\label{eq-08}
  Q_w = \left\{ \begin{array}{ll}
                -4.4(E_y-0.49)\left(P_d/3\right)^{\gamma} &  E_y > 0.49, \\
                0 & E_y \le 0.49 ,
              \end{array} \right.
\end{equation}
Where $E_y$ is the solar wind electric field, and $\gamma$ is a constant related to the solar wind dynamic pressure. In WCL equation, $\gamma$= 0.2. To see the role of dynamic pressure in different storms, here for each GGS event, we artificially adjust $\gamma$ so that the intensity of the geomagnetic storm estimated by WCL equation is closest to the real intensity of the storm.

For OM equation, we use $\Delta_{\textrm{om}}$ to indicate the difference between $\Delta$SYM-H$_{\textrm{ob}}^*$ and $\Delta$SYM-H$_{\textrm{om}}^*$, namely that $\Delta_{\textrm{om}} = \Delta \textrm{SYM-H}_{\textrm{ob}}^* - \Delta \textrm{SYM-H}_{\textrm{om}}^*$. For Burton equation, we use $\Delta_{b}$ to indicate the difference between $\Delta$SYM-H$_{ob}^*$ and $\Delta$SYM-H$_{b}^*$, namely that $\Delta_{b} = \Delta \textrm{SYM-H}_{ob}^* - \Delta \textrm{SYM-H}_{b}^*$. For WCL equation, $\Delta_{w} = \Delta \textrm{SYM-H}_{\textrm{ob}}^* - \Delta \textrm{SYM-H}_{\textrm{w}}^*$.

\section{Results} \label{sec-results}

\begin{figure}[ht!]
\plotone{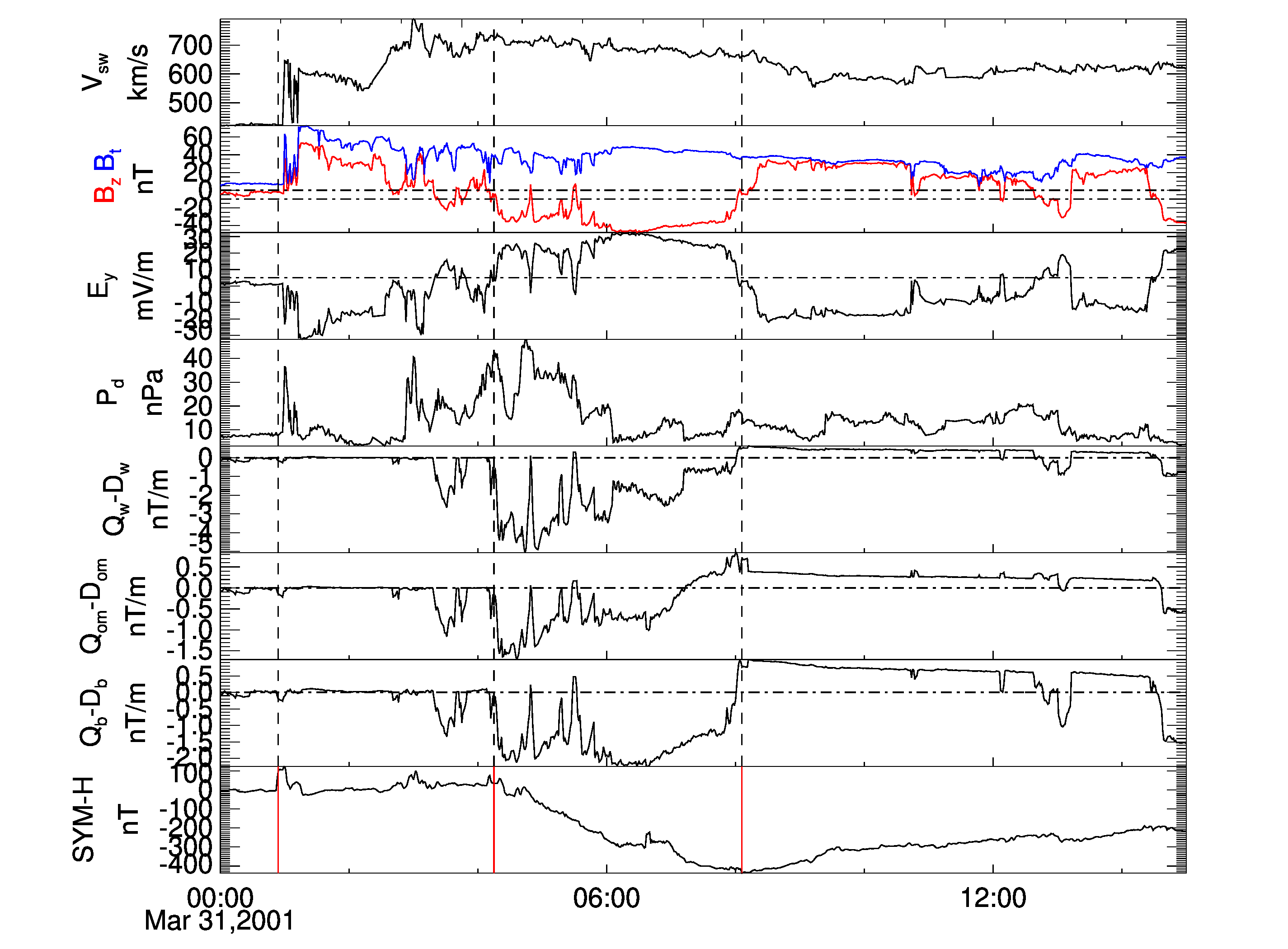}
\caption{The observations of solar wind parameters and the derived $Q-D$ for different empirical formulae and SYM-H index on 31 March 2001. From top to bottom, it shows solar wind speed ($V_{sw}$), interplanetary magnetic field (IMF) with blue line for the strength of IMF ($B_t$) and red line for the southward component of IMF, solar wind electric field ($E_y$), solar wind dynamic pressure ($P_d$), the derived $Q_w-D_w$ when $\gamma$=0.44, $Q_{om}-D_{om}$ and $Q_b-D_b$, SYM-H index, respectively. The horizontal dot dashed lines in the second panel indicate 0 and 10 nT, respectively. The horizontal dot dashed line in the third panel indicates $E_y$ = 5 $mV/m$.\label{fig-01}}
\end{figure}

Firstly, two examples are presented here for analysis. The first example is the storm that occurred on 31 March 2001 shown in Figure \ref{fig-01}. An interplanetary shock reached the magnetosphere at 00:52 UT on 31 March 2001 indicated by the first vertical red solid line in Figure \ref{fig-01}. The main phase of the storm is the period between the second and third red vertical solid lines. The interplanetary source of the storm main phase is the solar wind between the second and third vertical dashed lines. According to equation \ref{eq-07}, $\Delta$SYM-H$_{ob}^*$ is -489.5 nT. $\gamma$ is set as 0.44 in this case and the derived $Q_w-D_w$, $Q_{om}-D_{om}$ and $Q_b-D_b$ are shown in the 5th, 6th and 7th panels of Figure \ref{fig-01}, respectively. According to equation \ref{eq-03}, the derived $\Delta$SYM-H$_b^*$ and $\Delta$SYM-H$_{om}^*$ are -342.6 nT and -112.1 nT, respectively, while $\Delta$SYM-H$_w^*$ is -490.6 nT when $\gamma$ is equal to 0.44. $\Delta$SYM-H$_w^*$ is very close to $\Delta$SYM-H$_{ob}^*$. $\Delta_{om}$ and $\Delta_b$ are -377.4 nT and -147 nT, respectively.

\begin{figure}[ht!]
\plotone{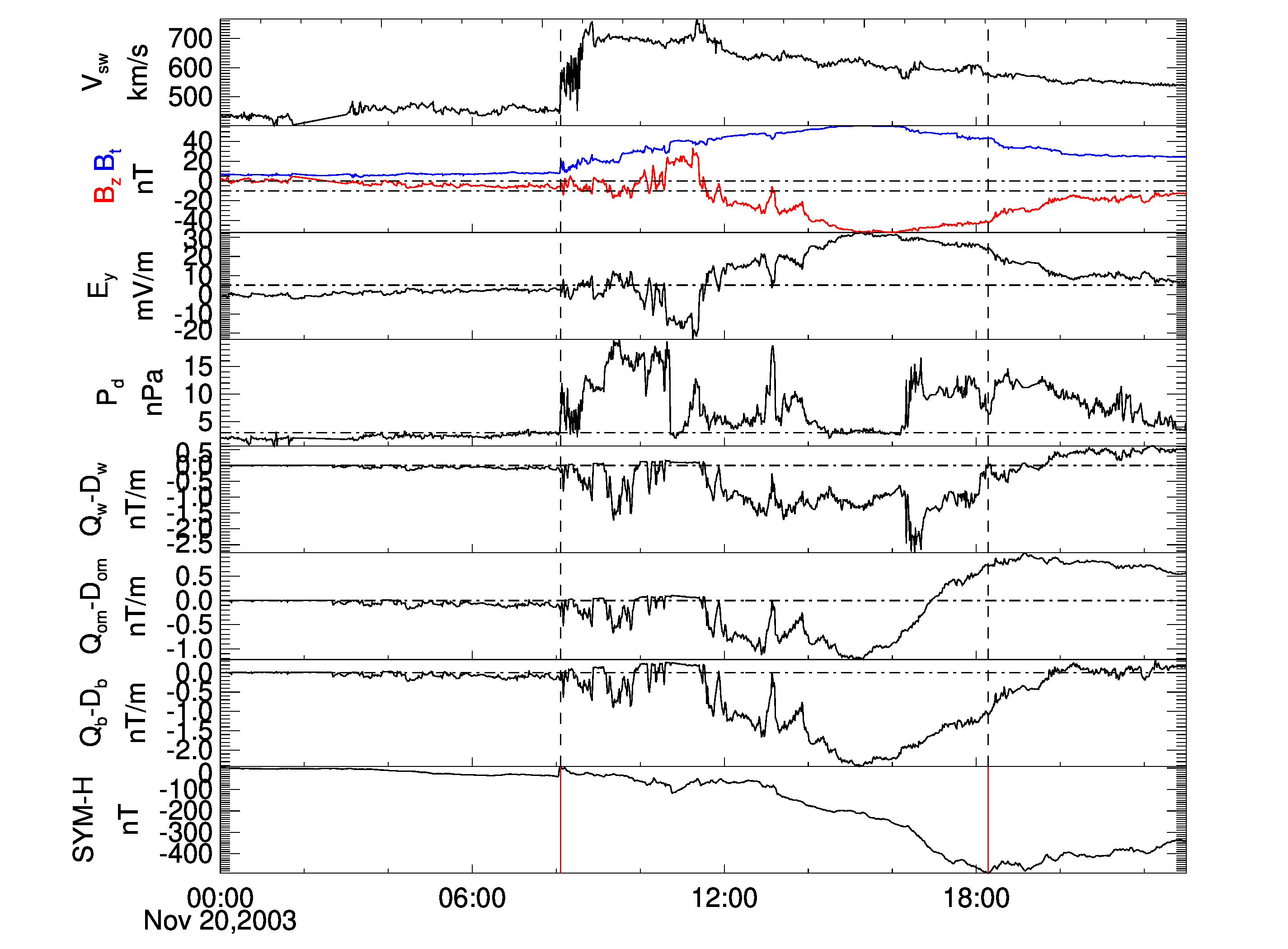}
\caption{The observations of solar wind parameters and the derived $Q-D$ for different empirical formulae and SYM-H index on 20 November 2003.\label{fig-02}}
\end{figure}

The second example is the storm that occurred on 20 November 2003. An interplanetary shock reached the magnetosphere at 08:02 UT on 20 November 2003 indicated by first vertical red solid line shown in Figure \ref{fig-02}. The main phase of the storm is the period between the first and second vertical red solid lines. The interplanetary cause of the storm main phase is the solar wind between the first and second vertical dashed lines. According to equation \ref{eq-07}, $\Delta$SYM-H$_{ob}^*$ is -505.4 nT. $\gamma$ is set as 0.44 in this case and the derived $Q_w-D_w$, $Q_{om}-D_{om}$ and $Q_b-D_b$ are shown in the 5th, 6th and 7th panels of Figure \ref{fig-02}, respectively. According to equation \ref{eq-03}, the derived $\Delta$SYM-H$_{b}^*$ and $\Delta$SYM-H$_{om}^*$ are -642.9 nT and -229.5 nT, respectively, and the derived $\Delta$SYM-H$_{w}^*$ is -507.1 nT when $\gamma$ is equal to 0.44. Note that $\Delta$SYM-H$_{w}^*$ is very close to $\Delta$SYM-H$_{ob}^*$, and $\Delta_{om}$ and $\Delta_b$ are -276 nT and 137.5 nT, respectively.

The intensities of 15 GGSs during solar cycle 23 are calculated by Burton equation, OM equation and WCL equation, respectively, and then compared with the observed intensities of the GGSs, as shown in Table \ref{tab-01}. In the Table, GGSs are numbered in column 1; the date that the GGS occurred in column 2; $\Delta$SYM-H$_{ob}^*$ in column 3; the interplanetary(IP) source responsible for the main phases of the GGSs in column 4; $\gamma$ used in WCL equation in column 5; the derived $\Delta$SYM-H$_{w}^*$ in column 6; $\Delta$SYM-H$_{om}^*$ in column 7; $\Delta_{om}$ in column 8; $\Delta_{om}/{\Delta \textrm{SYM-H}_{ob}^*}$ in column 9; $\Delta$SYM-H$_b^*$ in column 10; $\Delta_b$ in column 11; and $\Delta_b/{\Delta\textrm{SYM-H}_{ob}^*}$ in column 12.

\begin{figure}[ht!]
\plotone{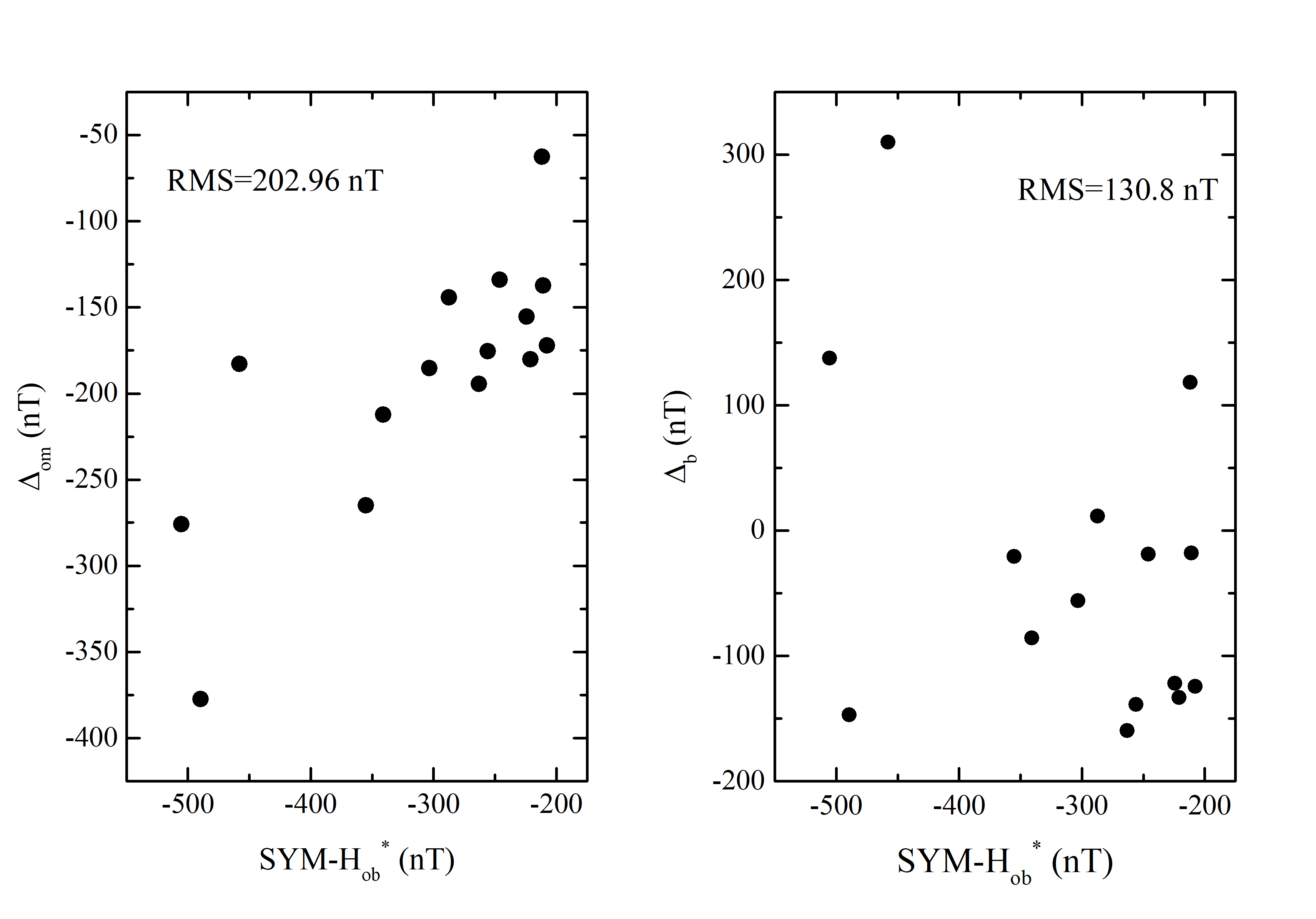}
\caption{The distribution of error with $\Delta$SYM-H$_{ob}^*$. The left panel is the distribution of $\Delta_{om}$ with $\Delta$SYM-H$_{ob}^*$, while the right panel is the distribution of $\Delta_b$ with $\Delta$SYM-H$_{ob}^*$.\label{fig-03}}
\end{figure}

According to $\Delta_{om}$, $\Delta_b$, and $\Delta$SYM-H$_{ob}^*$ shown in Table \ref{tab-01}, the distribution of $\Delta_{om}$ with $\Delta$SYM-H$_{ob}^*$ is shown in the left panel of Figure \ref{fig-03}, while the distribution of $\Delta_{b}$ with $\Delta$SYM-H$_{ob}^*$ is shown in the right panel of Figure \ref{fig-03}. The derived RMS (Root Mean Square) error between $\Delta$SYM-H$_{b}^*$ and $\Delta$SYM-H$_{ob}^*$ for 15 GGSs is 130.4 nT, while the RMS error between $\Delta$SYM-H$_{om}^*$ and $\Delta$SYM-H$_{ob}^*$ for 15 GGSs is 203 nT. The statistical results shown in Figure \ref{fig-03} prove that Burton equation is more accurate than OM equation in the estimation of GGSs intensities. As shown in Table \ref{tab-01}, the relative error for the storms with intensities $\Delta$SYM-H$<-$400 nT estimated by Burton equation is larger than 27\%, indicating that the intensity of an extreme storm estimated by Burton equation tend to have a large absolute error.

\begin{longrotatetable}
\begin{deluxetable*}{llllllllllll}
\tablecaption{The intensities of GGSs estimated by OM, Burton and WCL equations and the observed intensities of the GGSs\label{tab-01}}
\tablewidth{700pt}
\tabletypesize{\scriptsize}
\tablehead{
\colhead{} & \colhead{} &
\colhead{} & \colhead{} &
\multicolumn{2}{|c}{WCL Equation} &
\multicolumn{3}{|c}{OM Equation} &
\multicolumn{3}{|c|}{Burton Equation} \\
\cline{5-6}
\cline{7-9}
\cline{10-12}
\colhead{No.} & \colhead{Date} &
\colhead{$\Delta$SYM-H$_{ob}^*$} & \colhead{IP source} &
\colhead{$\gamma$} & \colhead{$\Delta$SYM-H$_w^*$} & \colhead{$\Delta$SYM-H$_{om}^*$} &
\colhead{$\Delta_{om}$} & \colhead{$\Delta_{}om/{\Delta\textrm{SYM-H}_{ob}^*}$} &
\colhead{$\Delta$SYM-H$_b^*$} & \colhead{$\Delta_b$} &
\colhead{$\Delta_b/{\Delta\textrm{SYM-H}_{ob}^*}$} \\
\colhead{} & \colhead{yyyy/mm/dd} & \colhead{(nT)} & \colhead{} &
\colhead{} & \colhead{(nT)} &
\colhead{(nT)} & \colhead{(nT)} & \colhead{} &
\colhead{(nT)} & \colhead{(nT)} & \colhead{}
}
\startdata
1  & 1998/05/04 & $-$246.1 & PICME-SH      & 1.05 & $-$250.8  & $-$112.1  & $-$134 & 54.4\% & $-$227.2 & $-$18.9  & 7.7\% \\
2  & 1998/09/25 & $-$210.8 & SH+{\bf MC}   & 0.82 & $-$210.7  & $-$73.4   & $-$137.4 & 65.2\% & $-$193 & $-$17.8  & 8.5\% \\
3  & 1999/09/22 & $-$224.3 & MC            & 0.49 & $-$225.7  & $-$68.8   & $-$155.5 & 69.3\% & $-$102.4 & $-$122 & 54.4\% \\
4  & 1999/10/22 & $-$287.3 & ICME - CIR    & 1.02 & $-$286.7 & $-$143.2  & $-$144.2 & 50.2\% & $-$298.6 & 11.3     & 3.9\% \\
5  & 2000/04/06 & $-$355.1 & SH            & 0.46 & $-$358.2  & $-$90.1   & $-$265    & 74.6\% & $-$334.4 & $-$20.7  & 5.8\% \\
6  & 2000/08/12 & $-$211.8 & SH+{\bf MC}   & 0.43 & $-$212.5  & $-$149.1  & $-$62.7  & 29.6\% & $-$330.1 & 118.4    & 55.9\% \\
7  & 2000/09/17 & $-$255.9 & {\bf SH(M)}+ICME    & 0.43 & $-$255.7  & $-$80.5   & $-$175.4 & 68.6\% & $-$117.2 & $-$138.7 & 54.2\% \\
8  & 2001/03/31 & $-$489.5 & {\bf SH}+MC   & 0.44 & $-$490.6  & $-$112.1  & $-$377.4 & 77.1\% & $-$342.6 & $-$147 & 30.0\% \\
9  & 2001/04/11 & $-$303.4 & {\bf SH}+MC   & 0.27 & $-$301.7  & $-$118.1  & $-$185.2 & 61.1\% & $-$247.5 & $-$56   & 18.4\% \\
10 & 2001/10/21 & $-$263.1 & {\bf SH}+MC   & 0.59 & $-$264.9  & $-$68.8 & $-$194.3 & 73.9\% & $-$103.6 & $-$159.5 & 60.6\% \\
11 & 2003/11/20 & $-$505.4 & SH+{\bf MC}   & 0.44 & $-$507.2  & $-$229.5 & $-$276. & 54.6\% & $-$642.9 & 137.5     & 27.2\% \\
12 & 2004/11/08 & $-$457.9 & SH+{\bf MC}   & 0.28 & $-$456.  & $-$275.1 & $-$182.8 & 39.9\% & $-$767.7 & 309.8    & 67.7\% \\
13 & 2004/11/09 & $-$207.7 & PICME-SH + MC & 0.50 & $-$209.7  & $-$35.6   & $-$172.1 & 82.9\% & $-$83.3  & $-$124.5 & 59.9\% \\
14 & 2005/05/15 & $-$340.9 & SH + MC       & 0.35 & $-$341  & $-$128.7  & $-$212.2 & 62.3\% & $-$255.2 & $-$85.7  & 25.1\% \\
15 & 2006/12/15 & $-$221. & SH+{\bf MC}     & 1.31 & $-$219.8  & $-$40.9  & $-$180.1 & 81.5\% & $-$87.7  & $-$133.3  & 60.3\% \\
\enddata
\tablecomments{The bold words indicate the structure that made a key contribution to the corresponding storm.\\
PICME-SH and PMC-SH denote a shock propagating through a preceding ICME or magnetic cloud respectively \citep{Zhang2007a}.\\
ICME-CIR indicates that both ICME and CIR contributed to the main phase of the storm \citep{Zhang2007a}.}
\end{deluxetable*}
\end{longrotatetable}

\section{Discussion and Summary} \label{sec-discussion}

It is usually accepted that the intensity of a geomagnetic storm completely depends on the solar wind electric field \citep[e.g.,][]{Burton1975, OBrien2000, WangYM2003} with solar wind dynamic pressure playing little role. This may be the reason why Burton equation and OM equation have been extensively used to estimate the intensities of extreme storms \citep[e.g.,][]{Tsurutani2003, Liu2014}, or why researchers only mentioned the solar wind electric field when talking about extreme storms \citep[e.g.,][]{Kumar2015, Balan2017}. However, $\Delta$SYM-H$_{om}^*$ is always smaller than $\Delta$SYM-H$_{ob}^*$ and most of $\Delta$SYM-H$_{om}^*$ are much smaller than $\Delta$SYM-H$_{ob}^*$ for 15 GGSs shown in Table \ref{tab-01}. The RMS error between $\Delta$SYM-H$_{ob}^*$ and $\Delta$SYM-H$_{\textrm{om}}^*$ is 203 nT, implying that if we use OM equation to estimate the intensity of an extreme geomagnetic storm, the estimation may be much smaller than the real intensity of the storm.

As shown in Figure \ref{fig-03} and Table \ref{tab-01}, the intensity of a GGS may be overestimated or underestimated by Burton equation. In general, Burton equation is more accurate than OM equation. However, $\Delta_b$ for the storms on 31 March 2001, 20 November 2003, and 8 November 2004 are -147 nT, 137.5 nT and 309.8 nT, respectively. The storm on 31 March 2001 is mainly caused by the sheath \citep{Cheng2020}, while the main phase of the storm on 20 November 2003 was mainly caused by the corresponding MCs with the sheaths making small contribution. This implies that a storm mainly caused by a sheath may be underestimated by Burton equation, while the intensity of a storm caused by a MC may be overestimated by Burton equation. The RMS error for the three storm with $\Delta$SYM-H$_{ob}^*$ $< -$400 nT between the estimated intensity and the observed intensity is 213.3 nT, implying that the averaged absolute error is very large if we use Burton equation to estimate the intensities of the storm with $\Delta$SYM-H$_{ob}^*$ $< -$400 nT.

\begin{figure}[ht!]
\plotone{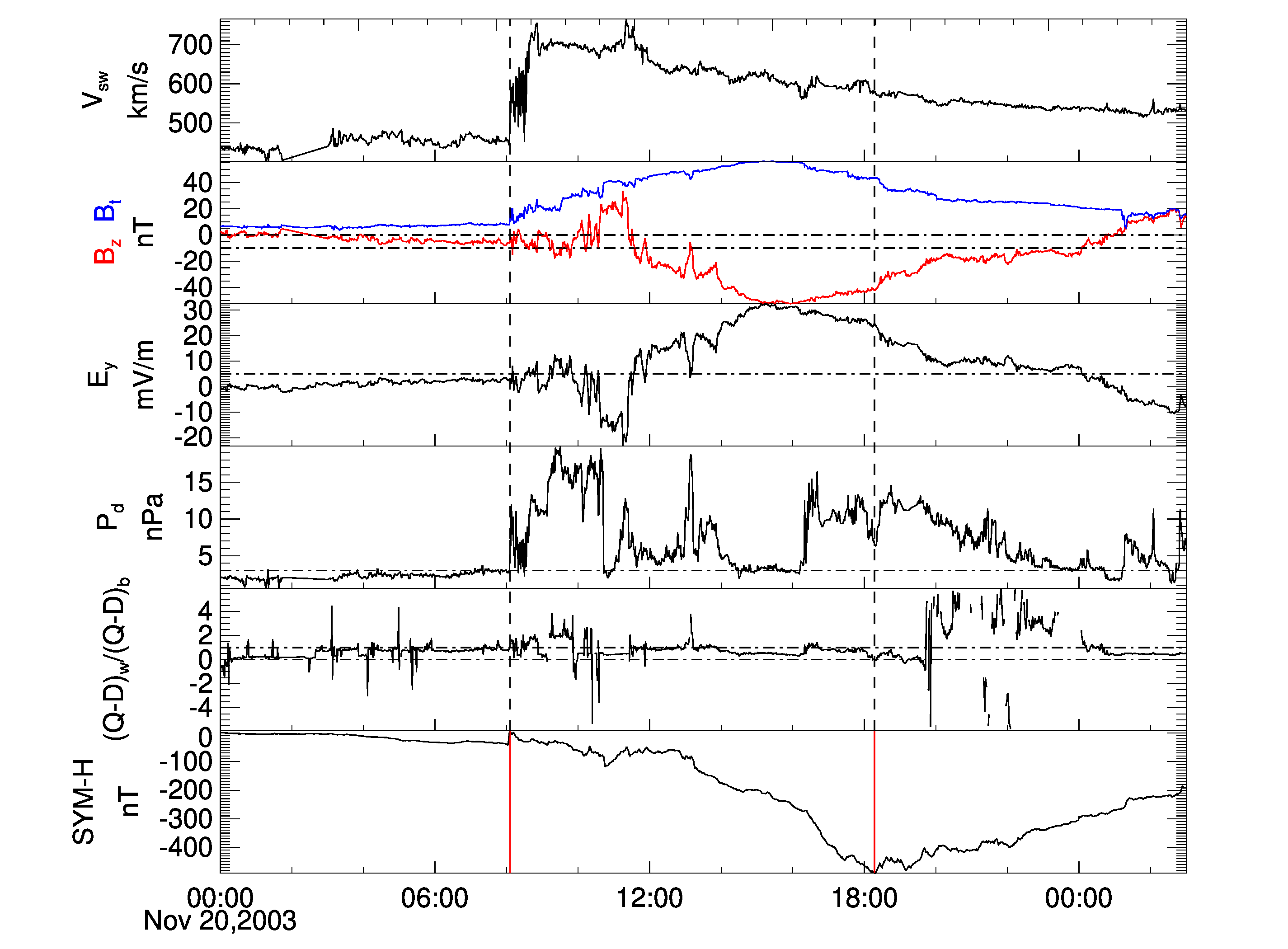}
\caption{The observations of solar wind parameters, the derived $(Q-D)_w/(Q-D)_b$ and SYM-H index on 20 November 2003. From top to bottom, it shows solar wind speed ($V_{sw}$), interplanetary magnetic field (IMF) with blue line for the strength of IMF ($B_t$) and red line for the southward component of IMF, solar wind electric field ($E_y$), solar wind dynamic pressure ($P_d$), the derived $(Q-D)_w/(Q-D)_b$ when $\gamma$=0.43, \textbf{and} SYM-H index, respectively. The horizontal dot dashed line in the fifth panel indicates $(Q-D)_w/(Q-D)_b$ = 1.\label{fig-04}}
\end{figure}

Why is the intensity of the storm on 20 November 2003 estimated by Burton equation stronger than the real intensity of the storm? Figure \ref{fig-04} gives the answer to this question. We use $(Q-D)_w/(Q-D)_b$ to indicate $(Q_w-D_w)/(Q_b-D_b)$ and the derived $(Q-D)_w/(Q-D)_b$ is shown in 5th panel of Figure \ref{fig-04}. We can see from Figure \ref{fig-04} that $(Q-D)_w/(Q-D)_b < 1$ during the period around the maximum of $E_y$ when solar wind dynamic pressure is lower than 3 nPa. This is the reason why the estimated intensity of the storm is stronger than the observed intensity of the storm.

Sheath compression regions usually have higher speed and higher dynamic pressure and therefore have higher efficiency in the generation of magnetic storms than MCs \citep{Yermolaev2010, Yermolaev2012}, implying that solar wind dynamic pressure is an important parameter for the development of a geomagnetic storm. \cite{Newell2007} found that Dst did not correlate best with $d\Phi_{MP}/dt$, where $d\Phi_{MP}/dt = V_{sw}^{4/3}B_T^{2/3}sin^{8/3}(arctan(B_y/B_z)/2)$. However, Dst has a good correlation with $P^{1/2}d\Phi_{MP}/dt$, suggesting that solar wind dynamic pressure is an important factor for the intensity of a geomagnetic storm. Case studies \citep[e.g.,][]{Kataoka2005, Cheng2020, Liu2021} and statistical studies \citep[e.g.,][]{Le2020, Zhao2021} provide robust evidence that the solar wind dynamic pressure is also an important parameter for the intensity of a geomagnetic storm beside the solar wind electric field. Case study \citep{Cheng2020, Liu2021} proved that the empirical formula built by \cite{WangYM2003} was not correct because it ignored the contribution of solar wind dynamic pressure to the intensity of a geomagnetic storm made by solar wind dynamic pressure. The empirical formula proposed by \cite{Gopalswamy2018}, which only relates the intensity of a geomagnetic storm to the largest value of $|V_{sw}B_s|$, cannot reflect the sustained interaction between solar wind and the magnetosphere during the main phase of a geomagnetic storm. In addition, the empirical formula proposed by \cite{Gopalswamy2018} did not consider the contribution to the geomagnetic storm intensity made by solar wind dynamic pressure. \textbf{In this context, the empirical formula proposed by \cite{Gopalswamy2018} may be incomplete.}

The results of the present study indicate that Burton equation and OM equation can not work effectively when the two equations are used to estimate the intensities of GGSs. On the contrary, WCL equation is a very good empirical formulae. For each GGS, we can always find a right $\gamma$, so that the intensity of the GGS estimated by WCL equation can be as close as possible to the observed intensity of the storm.

The interaction process between the solar wind and the magnetosphere to produce a storm is very complicated, and it varies from one storm to another. This may be the reason why we cannot always use a unified $\gamma$ in the WCL equation to estimate the intensities of different geomagnetic storms, namely that different storms may need different $\gamma$ so as to ensure that the intensity of a GGS estimated by WCL equation can always be very close to the observed intensity of the storm. This may also be the reason why it is difficult to build an ideal empirical formula relating the intensity of a geomagnetic storm to the solar wind parameters precisely. \textbf{Anyway, the results of present study also prove that the solar wind dynamic pressure is an important factor for the intensity of a GGS, namely that solar wind dynamic pressure plays an important and complex role in the solar wind-magnetosphere coupling. However, the contribution made by solar wind dynamic pressure to the intensity of a GGS is ignored by Burton equation and OM equation because the ring current injection terms in Burton equation and OM equation have no solar wind dynamic pressure. This is the reason why Burton equation and OM equation cannot work effectively when the two equations are used to estimate the intensities of GGSs.}.

This study focuses on the evaluation of Burton equation, OM equation and WCL equations. The empirical formulae \citep[e.g.,][]{Fenrich1998, Temerin2002, Ballatore2003, Temerin2006, Boynton2011} will be evaluated in the near future.

\begin{acknowledgments}
We thank NASA for providing the solar wind data and the Center for Geomagnetism and Space Magnetism, Kyoto University, for providing the SYM-H index. We also thank Institute of Geophysics, China Earthquake Administration for providing sudden storm commence data. This work is jointly supported by Sino-South Africa Joint Research on Polar Space Environment (2021YFE0106400), International Cooperation Project on Scientific and Technological Innovation Between Governments, National key Plans on Research and Development, Ministry of Science and Technology, China, CAS Key Laboratory of Solar Activity under number KLSA (grant No. KLSA202109), and the National Natural Science Foundation of China (Grant Nos. 41074132, 41274193, 41474166, 41774195 and 41874187).
\end{acknowledgments}

\bibliography{GGS2021}{}

\begin{thebibliography}{}
\expandafter\ifx\csname natexlab\endcsname\relax\def\natexlab#1{#1}\fi
\providecommand{\url}[1]{\href{#1}{#1}}
\providecommand{\dodoi}[1]{doi:~\href{http://doi.org/#1}{\nolinkurl{#1}}}
\providecommand{\doeprint}[1]{\href{http://ascl.net/#1}{\nolinkurl{http://ascl.net/#1}}}
\providecommand{\doarXiv}[1]{\href{https://arxiv.org/abs/#1}{\nolinkurl{https://arxiv.org/abs/#1}}}

\bibitem[{Akasofu(1981)}]{Akasofu1981}
Akasofu, S.~I. 1981, Space Science Reviews, 28, 121, \dodoi{10.1007/BF00218810}

\bibitem[{Arowolo {et~al.}(2021)Arowolo, Akala, \& Oyeyemi}]{Arowolo2021}
Arowolo, O.~A., Akala, A.~O., \& Oyeyemi, E.~O. 2021, Journal of Geophysical
  Research: Space Physics, 126, e2020JA027929,
  \dodoi{https://doi.org/10.1029/2020JA027929}

\bibitem[{Balan {et~al.}(2017)Balan, Ebihara, Skoug, Shiokawa, Batista,
  Tulasi~Ram, Omura, Nakamura, \& Fok}]{Balan2017}
Balan, N., Ebihara, Y., Skoug, R., {et~al.} 2017, Journal of Geophysical
  Research: Space Physics, 122, 2824,
  \dodoi{https://doi.org/10.1002/2016JA023853}

\bibitem[{Ballatore \& Gonzalez(2003)}]{Ballatore2003}
Ballatore, P., \& Gonzalez, W.~D. 2003, Earth, Planets and Space, 55, 427,
  \dodoi{10.1186/BF03351776}

\bibitem[{{Bhaskar, Ankush} \& {Vichare, Geeta}(2019)}]{Bhaskar2019}
{Bhaskar, Ankush}, \& {Vichare, Geeta}. 2019, J. Space Weather Space Clim., 9,
  A12, \dodoi{10.1051/swsc/2019007}

\bibitem[{Borovsky(2014)}]{Borovsky2014}
Borovsky, J.~E. 2014, Journal of Geophysical Research: Space Physics, 119,
  5364, \dodoi{https://doi.org/10.1002/2013JA019607}

\bibitem[{Borovsky \& Denton(2006)}]{Borovsky2006}
Borovsky, J.~E., \& Denton, M.~H. 2006, Journal of Geophysical Research: Space
  Physics, 111, \dodoi{https://doi.org/10.1029/2005JA011447}

\bibitem[{Boynton {et~al.}(2011)Boynton, Balikhin, Billings, Sharma, \&
  Amariutei}]{Boynton2011}
Boynton, R.~J., Balikhin, M.~A., Billings, S.~A., Sharma, A.~S., \& Amariutei,
  O.~A. 2011, Annales Geophysicae, 29, 965, \dodoi{10.5194/angeo-29-965-2011}

\bibitem[{Burton {et~al.}(1975)Burton, McPherron, \& Russell}]{Burton1975}
Burton, R.~K., McPherron, R.~L., \& Russell, C.~T. 1975, Journal of Geophysical
  Research (1896-1977), 80, 4204,
  \dodoi{https://doi.org/10.1029/JA080i031p04204}

\bibitem[{Cheng {et~al.}(2020)Cheng, Le, \& Zhao}]{Cheng2020}
Cheng, L.-B., Le, G.-M., \& Zhao, M.-X. 2020, Research in Astronomy and
  Astrophysics, 20, 036, \dodoi{10.1088/1674-4527/20/3/36}

\bibitem[{Council(2008)}]{Baker2008}
Council, N.~R. 2008, Severe Space Weather Events: Understanding Societal and
  Economic Impacts: A Workshop Report (Washington, DC: The National Academies
  Press), \dodoi{10.17226/12507}

\bibitem[{Dal~Lago {et~al.}(2006)Dal~Lago, Gonzalez, Balmaceda, Vieira, Echer,
  Guarnieri, Santos, da~Silva, de~Lucas, Clua~de Gonzalez, Schwenn, \&
  Schuch}]{Dal2006}
Dal~Lago, A., Gonzalez, W.~D., Balmaceda, L.~A., {et~al.} 2006, Journal of
  Geophysical Research: Space Physics, 111,
  \dodoi{https://doi.org/10.1029/2005JA011394}

\bibitem[{Du {et~al.}(2008)Du, Tsurutani, \& Sun}]{Du2008}
Du, A.~M., Tsurutani, B.~T., \& Sun, W. 2008, Journal of Geophysical Research:
  Space Physics, 113, \dodoi{https://doi.org/10.1029/2008JA013284}

\bibitem[{Eastwood {et~al.}(2017)Eastwood, Biffis, Hapgood, Green, Bisi,
  Bentley, Wicks, McKinnell, Gibbs, \& Burnett}]{Eastwood2017}
Eastwood, J.~P., Biffis, E., Hapgood, M.~A., {et~al.} 2017, Risk Analysis, 37,
  206, \dodoi{https://doi.org/10.1111/risa.12765}

\bibitem[{Fenrich \& Luhmann(1998)}]{Fenrich1998}
Fenrich, F.~R., \& Luhmann, J.~G. 1998, Geophysical Research Letters, 25, 2999,
  \dodoi{https://doi.org/10.1029/98GL51180}

\bibitem[{Gopalswamy(2018)}]{Gopalswamy2018}
Gopalswamy, N. 2018, in Extreme Events in Geospace, ed. N.~Buzulukova
  (Elsevier), 37--63,
  \dodoi{https://doi.org/10.1016/B978-0-12-812700-1.00002-9}

\bibitem[{Gosling {et~al.}(1991)Gosling, McComas, Phillips, \&
  Bame}]{Gosling1991}
Gosling, J.~T., McComas, D.~J., Phillips, J.~L., \& Bame, S.~J. 1991, Journal
  of Geophysical Research: Space Physics, 96, 7831,
  \dodoi{https://doi.org/10.1029/91JA00316}

\bibitem[{Grandin {et~al.}(2019)Grandin, Aikio, \& Kozlovsky}]{Grandin2019}
Grandin, M., Aikio, A.~T., \& Kozlovsky, A. 2019, Journal of Geophysical
  Research: Space Physics, 124, 3871,
  \dodoi{https://doi.org/10.1029/2018JA026396}

\bibitem[{Hajra {et~al.}(2020)Hajra, Tsurutani, \& Lakhina}]{Hajra2020}
Hajra, R., Tsurutani, B.~T., \& Lakhina, G.~S. 2020, The Astrophysical Journal,
  899, 3, \dodoi{10.3847/1538-4357/aba2c5}

\bibitem[{Iyemori(1990)}]{Iyemori1990}
Iyemori, T. 1990, Journal of geomagnetism and geoelectricity, 42, 1249,
  \dodoi{10.5636/jgg.42.1249}

\bibitem[{Iyemori {et~al.}(2010)Iyemori, Takeda, Nose, Odagi, \&
  Toh}]{Iyemori2010}
Iyemori, T., Takeda, M., Nose, M., Odagi, Y., \& Toh, H. 2010, {Mid-latitude}
  {Geomagnetic} {Indices} {ASY} and {SYM} for 2009 {(Provisional)} (Kyoto
  University, Japan: Internal Report of Data Analysis Center for Geomagnetism
  and Space Magnetism).
\newblock
  \url{http://isgi.unistra.fr/Documents/References/Iyemori_et_al_2010.pdf}

\bibitem[{Kan \& Lee(1979)}]{Kan1979}
Kan, J.~R., \& Lee, L.~C. 1979, Geophysical Research Letters, 6, 577,
  \dodoi{https://doi.org/10.1029/GL006i007p00577}

\bibitem[{Kataoka {et~al.}(2005)Kataoka, Fairfield, Sibeck, Rastätter, Fok,
  Nagatsuma, \& Ebihara}]{Kataoka2005}
Kataoka, R., Fairfield, D.~H., Sibeck, D.~G., {et~al.} 2005, Geophysical
  Research Letters, 32, \dodoi{https://doi.org/10.1029/2005GL024495}

\bibitem[{Kozyra \& Liemohn(2003)}]{Kozyra2003}
Kozyra, J.~U., \& Liemohn, M.~W. 2003, Space Science Reviews, 109, 105,
  \dodoi{10.1023/B:SPAC.0000007516.10433.ad}

\bibitem[{Kumar {et~al.}(2015)Kumar, Veenadhari, Tulasi~Ram, Selvakumaran,
  Mukherjee, Singh, \& Kadam}]{Kumar2015}
Kumar, S., Veenadhari, B., Tulasi~Ram, S., {et~al.} 2015, Journal of
  Geophysical Research: Space Physics, 120, 7307,
  \dodoi{https://doi.org/10.1002/2015JA021661}

\bibitem[{Le \& Liu(2020)}]{Le2020Sol2952}
Le, G.-M., \& Liu, G.-A. 2020, Solar Physics, 295, 35,
  \dodoi{10.1007/s11207-020-01600-8}

\bibitem[{Le {et~al.}(2020)Le, Liu, \& Zhao}]{Le2020}
Le, G.-M., Liu, G.-A., \& Zhao, M.-X. 2020, Solar Physics, 295, 108,
  \dodoi{10.1007/s11207-020-01675-3}

\bibitem[{Le {et~al.}(2021{\natexlab{a}})Le, Zhang, \& Zhao}]{Le2021Sol2961}
Le, G.-M., Zhang, Y.-N., \& Zhao, M.-X. 2021{\natexlab{a}}, Solar Physics, 296,
  16, \dodoi{10.1007/s11207-020-01758-1}

\bibitem[{Le {et~al.}(2021{\natexlab{b}})Le, Zhao, ~, Liu, Mao, \&
  Xu}]{Le2021MNRAS}
Le, G.-M., Zhao, M.-X., ~, Q.-L., {et~al.} 2021{\natexlab{b}}, Monthly Notices
  of the Royal Astronomical Society, 502, 2043, \dodoi{10.1093/mnras/stab169}

\bibitem[{Le {et~al.}(2021{\natexlab{c}})Le, Zhao, Zhang, \&
  Liu}]{Le2021Solinpress}
Le, G.-M., Zhao, M.-X., Zhang, W.-T., \& Liu, G.-A. 2021{\natexlab{c}}, Solar
  Physics, 296, 187, \dodoi{10.1007/s11207-021-01927-w}

\bibitem[{Le {et~al.}(2016)Le, Li, Tang, Ding, Yin, Chen, Lu, Chen, \&
  Li}]{Le2016}
Le, G.-M., Li, C., Tang, Y.-H., {et~al.} 2016, Research in Astronomy and
  Astrophysics, 16, 014, \dodoi{10.1088/1674-4527/16/1/014}

\bibitem[{Liu {et~al.}(2021)Liu, Zhao, Le, \& Mao}]{Liu2021}
Liu, G.-A., Zhao, M.-X., Le, G.-M., \& Mao, T. 2021, Research in Astronomy and
  Astrophysics, \dodoi{10.1088/1674-4527/ac3126}

\bibitem[{Liu {et~al.}(2014)Liu, Luhmann, Kajdič, Kilpua, Lugaz, Nitta,
  Möstl, Lavraud, Bale, Farrugia, \& Galvin}]{Liu2014}
Liu, Y.~D., Luhmann, J.~G., Kajdič, P., {et~al.} 2014, Nature Communications,
  5, 3481, \dodoi{10.1038/ncomms4481}

\bibitem[{Love(2021)}]{Love2021}
Love, J.~J. 2021, Space Weather, 19, e2020SW002579,
  \dodoi{https://doi.org/10.1029/2020SW002579}

\bibitem[{Miyoshi \& Kataoka(2005)}]{Miyoshi2005}
Miyoshi, Y., \& Kataoka, R. 2005, Geophysical Research Letters, 32,
  \dodoi{https://doi.org/10.1029/2005GL024590}

\bibitem[{Newell {et~al.}(2007)Newell, Sotirelis, Liou, Meng, \&
  Rich}]{Newell2007}
Newell, P.~T., Sotirelis, T., Liou, K., Meng, C.-I., \& Rich, F.~J. 2007,
  Journal of Geophysical Research: Space Physics, 112,
  \dodoi{https://doi.org/10.1029/2006JA012015}

\bibitem[{O'Brien \& McPherron(2000)}]{OBrien2000}
O'Brien, T.~P., \& McPherron, R.~L. 2000, Journal of Geophysical Research:
  Space Physics, 105, 7707, \dodoi{10.1029/1998JA000437}

\bibitem[{Pandya {et~al.}(2019)Pandya, Bhaskara, Ebihara, Kanekal, \&
  Baker}]{Pandya2019}
Pandya, M., Bhaskara, V., Ebihara, Y., Kanekal, S.~G., \& Baker, D.~N. 2019,
  Journal of Geophysical Research: Space Physics, 124, 6524,
  \dodoi{https://doi.org/10.1029/2019JA026771}

\bibitem[{Richardson {et~al.}(2002)Richardson, Cane, \&
  Cliver}]{Richardson2002}
Richardson, I.~G., Cane, H.~V., \& Cliver, E.~W. 2002, Journal of Geophysical
  Research: Space Physics, 107, SSH 8,
  \dodoi{https://doi.org/10.1029/2001JA000504}

\bibitem[{Riley \& Love(2017)}]{Riley2017}
Riley, P., \& Love, J.~J. 2017, Space Weather, 15, 53,
  \dodoi{https://doi.org/10.1002/2016SW001470}

\bibitem[{Sandhu {et~al.}(2021)Sandhu, Rae, \& Walach}]{Sandhu2021}
Sandhu, J.~K., Rae, I.~J., \& Walach, M.-T. 2021, Journal of Geophysical
  Research: Space Physics, 126, e2020JA028423,
  \dodoi{https://doi.org/10.1029/2020JA028423}

\bibitem[{Schulte in~den B\"aumen {et~al.}(2014)Schulte in~den B\"aumen, Moran,
  Lenzen, Cairns, \& Steenge}]{Schulte2014}
Schulte in~den B\"aumen, H., Moran, D., Lenzen, M., Cairns, I., \& Steenge, A.
  2014, Natural Hazards and Earth System Sciences, 14, 2749,
  \dodoi{10.5194/nhess-14-2749-2014}

\bibitem[{Shen {et~al.}(2017)Shen, Hudson, Jaynes, Shi, Tian, Claudepierre,
  Qin, Zong, \& Sun}]{Shen2017}
Shen, X.-C., Hudson, M.~K., Jaynes, A.~N., {et~al.} 2017, Journal of
  Geophysical Research: Space Physics, 122, 8327,
  \dodoi{https://doi.org/10.1002/2017JA024100}

\bibitem[{Temerin \& Li(2002)}]{Temerin2002}
Temerin, M., \& Li, X. 2002, Journal of Geophysical Research: Space Physics,
  107, SMP 31, \dodoi{https://doi.org/10.1029/2001JA007532}

\bibitem[{Temerin \& Li(2006)}]{Temerin2006}
---. 2006, Journal of Geophysical Research: Space Physics, 111,
  \dodoi{https://doi.org/10.1029/2005JA011257}

\bibitem[{Tenfjord \& Østgaard(2013)}]{Tenfjord2013}
Tenfjord, P., \& Østgaard, N. 2013, Journal of Geophysical Research: Space
  Physics, 118, 5659, \dodoi{https://doi.org/10.1002/jgra.50545}

\bibitem[{Tsurutani {et~al.}(2003)Tsurutani, Gonzalez, Lakhina, \&
  Alex}]{Tsurutani2003}
Tsurutani, B.~T., Gonzalez, W.~D., Lakhina, G.~S., \& Alex, S. 2003, Journal of
  Geophysical Research: Space Physics, 108,
  \dodoi{https://doi.org/10.1029/2002JA009504}

\bibitem[{Wang {et~al.}(2003{\natexlab{a}})Wang, Chao, \& Lin}]{WangCB2003}
Wang, C.~B., Chao, J.~K., \& Lin, C.-H. 2003{\natexlab{a}}, Journal of
  Geophysical Research: Space Physics, 108,
  \dodoi{https://doi.org/10.1029/2003JA009851}

\bibitem[{Wang {et~al.}(2003{\natexlab{b}})Wang, Shen, Wang, \&
  Ye}]{WangYM2003}
Wang, Y., Shen, C.~L., Wang, S., \& Ye, P.~Z. 2003{\natexlab{b}}, Geophysical
  Research Letters, 30, \dodoi{https://doi.org/10.1029/2003GL017901}

\bibitem[{Wanliss \& Showalter(2006)}]{Wanliss2006}
Wanliss, J.~A., \& Showalter, K.~M. 2006, Journal of Geophysical Research:
  Space Physics, 111, \dodoi{https://doi.org/10.1029/2005JA011034}

\bibitem[{Wygant {et~al.}(1983)Wygant, Torbert, \& Mozer}]{Wygant1983}
Wygant, J.~R., Torbert, R.~B., \& Mozer, F.~S. 1983, Journal of Geophysical
  Research: Space Physics, 88, 5727,
  \dodoi{https://doi.org/10.1029/JA088iA07p05727}

\bibitem[{Yermolaev {et~al.}(2021)Yermolaev, Lodkina, Dremukhina, Yermolaev, \&
  Khokhlachev}]{Yermolaev2021}
Yermolaev, Y.~I., Lodkina, I.~G., Dremukhina, L.~A., Yermolaev, M.~Y., \&
  Khokhlachev, A.~A. 2021, Universe, 7, \dodoi{10.3390/universe7050138}

\bibitem[{Yermolaev {et~al.}(2010)Yermolaev, Nikolaeva, Lodkina, \&
  Yermolaev}]{Yermolaev2010}
Yermolaev, Y.~I., Nikolaeva, N.~S., Lodkina, I.~G., \& Yermolaev, M.~Y. 2010,
  Annales Geophysicae, 28, 2177, \dodoi{10.5194/angeo-28-2177-2010}

\bibitem[{Yermolaev {et~al.}(2012)Yermolaev, Nikolaeva, Lodkina, \&
  Yermolaev}]{Yermolaev2012}
---. 2012, Journal of Geophysical Research: Space Physics, 117,
  \dodoi{https://doi.org/10.1029/2011JA017139}

\bibitem[{Zhang {et~al.}(2007)Zhang, Richardson, Webb, Gopalswamy, Huttunen,
  Kasper, Nitta, Poomvises, Thompson, Wu, Yashiro, \& Zhukov}]{Zhang2007a}
Zhang, J., Richardson, I.~G., Webb, D.~F., {et~al.} 2007, Journal of
  Geophysical Research: Space Physics, 112,
  \dodoi{https://doi.org/10.1029/2007JA012321}

\bibitem[{Zhao {et~al.}(2021)Zhao, Le, Li, Liu, \& Mao}]{Zhao2021}
Zhao, M.-X., Le, G.-M., Li, Q., Liu, G.-A., \& Mao, T. 2021, Solar Physics,
  296, 66, \dodoi{10.1007/s11207-021-01816-2}

\bibitem[{Zurbuchen \& Richardson(2006)}]{Zurbuchen2006}
Zurbuchen, T.~H., \& Richardson, I.~G. 2006, Space Science Reviews, 123, 31,
  \dodoi{10.1007/s11214-006-9010-4}

\end{thebibliography}
\bibliographystyle{aasjournal}

\end{document}